\documentclass[twocolumn,preprintnumbers,amssymb,prb,superscriptaddress]{revtex4}
\usepackage{graphicx}
\usepackage{dcolumn}
\usepackage{bm}
\usepackage{psfrag}
\begin{document}
\title{ On the statistics of resonances and
non-orthogonal eigenfunctions  in a 
model for single-channel chaotic scattering}
\author{Yan V. Fyodorov}
\affiliation{Department of Mathematical Sciences, Brunel
University, Uxbridge UB83PH, UK}
\affiliation{Petersburg Nuclear Physics Institute, Gatchina 188350, Russia}
\author{B. Mehlig}
\altaffiliation[On leave from: ]{School of Physics and Engineering Physics,
Chalmers/GU, 41296 G\"{o}teborg, Sweden }
\affiliation{Institute for Nuclear Theory, University of Washington,
Box 351550, Seattle  WA, 98195}

\begin{abstract}
We describe analytical and numerical results on the
statistical
properties of complex eigenvalues and the
corresponding non-orthogonal
eigenvectors for non-Hermitian random matrices
modeling one-channel quantum-chaotic scattering in 
systems with broken time-reversal invariance.
\end{abstract}
\maketitle
The statistical properties of non-orthogonal
eigenvectors
of large non-selfadjoint random matrices 
have recently been characterised in 
Refs.~\onlinecite{CM,Schom,Janik,Greb,MS}. 

Correlations of non-orthogonal eigenvectors 
are expected to determine dynamical properties of
classical random systems 
described by non-selfadjoint operators, such as
Fokker-Planck operators\cite{chalker}
 for example; they also play an important role in
quantum systems: 
in Ref.~\onlinecite{Schom} it was observed that the
statistics of 
non-orthogonal eigenvectors determines the properties 
of random lasing media. This has led 
to an increased interest in eigenvector statistics 
in non-selfadjoint random matrix ensembles (see also
Ref.~\onlinecite{Haake}).

In a model for quantum-chaotic scattering, 
the complex eigenvalues ${\cal E}_k\,,k=1,...,N$ 
of a random $N\!\times\!N$ non-Hermitian matrix (the
so-called ``effective Hamiltonian") 
${\cal H}_N=\hat{H}-i\hat{\Gamma}$ are used to
describe generic statistical properties  of 
resonances in quantum chaotic scattering (see
Ref.~\onlinecite{FS} and references therein):
for systems with broken time-reversal invariance (anti-unitary symmetry),
the matrices $\hat{H}$ are random $N\!\times\!N$
matrices from 
the Gaussian Unitary Ensemble~\cite{mehta}
with joint probability density $P(H)\,dH \propto
\exp[-(N/2) \mbox{Tr} H^2]\,dH$. 
In the limit of large
$N$, the mean eigenvalue density $\nu(E)$
for such matrices
is given by the semicircular law
$\nu(E)=(2\pi)^{-1}\sqrt{4-E^2}$
for $|E|<2$ (and zero otherwise). 
The corresponding mean spacing between neigbouring
eigenvalues around the point $E$ in the spectrum is
given by
$\Delta(E)=1/[N\nu(E)]$.

The Hermitian matrices $\hat{H}$ describe the
energy-level
statistics of the closed
counterpart of the scattering system; the 
Hermitian $N\!\times\!N$ matrix
$\hat{\Gamma}>0$ models the coupling of the system to
scattering
continua via $M=1,2,...$ open channels. It has rank
$M\le N$. For our purposes it can be chosen
diagonal:
$\hat{\Gamma}=\mbox{diag}(\gamma_1,\gamma_2,...,
\gamma_M,0,...,0)$.
The constants $0<\gamma_c<\infty$ parametrise the 
strength of the coupling 
to the scattering continua via a given channel
$c=1,...,M$. 
Here $\gamma_c=0$
corresponds to a closed channel $c$, and $\gamma_c=1$
describes  the so-called
perfectly coupled channel. 
Empirical situations correspond to the regime of large
$N$,
with $M$ fixed and $M\ll N$.  
Then the widths $\Gamma_k=2\,\mbox{Im}\,{\cal E}_k$ are of
the same order $1/N$
as the mean spacing $\Delta(E)$ between the positions
of the neighbouring 
resonances along the real energy axis. 
In this regime, the resonances
may partly or considerably overlap and first-order
perturbation theory
valid for small resonance overlaps breaks down.
Similarly,
self-consistent perturbation schemes\cite{CM,Janik,MS,chalker} assuming
many channels and strongly overlapping resonances
are inapplicable.

A detailed analytical understanding of
the statistical properties of the resonances in the
regime of partial
overlap has recently been achieved for the  case
of systems 
with broken time reversal invariance \cite{FS,FK}. 
These results, based on the random matrix approach, 
are expected to be applicable to a broad
class of quantum-chaotic systems. Indeed, 
the distribution of the widths $\Gamma_k$ derived in
Ref.~\onlinecite{FS}
 is in good agreement with available numerical data
for
quite diverse models of quantum chaotic
scattering\cite{Kottos,Kol}.   

Much less is known on properties of non-orthogonal
eigenvectors. Let $|R_k\rangle$ and $\langle
L_k|$ denote the right and the left eigenvectors of the
matrix $\hat{\cal H}$
corresponding to the eigenvalue ${\cal E}_k \equiv
E_k - i Y_k = E_k -
i\Gamma_k/2$,
\begin{eqnarray}
{\cal H}|R_k\rangle&=&{\cal E}_k|R_k\rangle\,,\quad
\langle L_k|{\cal H}\,\,\,=\langle L_k|{\cal E}_k\\ 
\nonumber
{\cal H}^{\dagger}|L_k\rangle&=& {\cal
E}^*_k|L_k\rangle\,,\quad 
\langle R_k|{\cal H}^{\dagger}=\langle R_k|{\cal
E}^*_k
\end{eqnarray} 
where the symbols $\dagger$ and $^*$ stand for
Hermitian
conjugation and complex conjugation, respectively.
Except for
a set of measure zero, the eigenvalues are
non-degenerate.
In this case the eigenvectors
form a complete, bi-orthogonal set. They can be
normalised 
to satisfy
$\langle L_k|R_l\rangle=\delta_{kl}$\,.
The most natural way to characterise the
non-orthogonality
of eigenvectors is to consider statistics of the
overlap matrix 
${\cal O}_{kl}=\langle L_k|L_l\rangle
\langle R_l|R_k\rangle$. 
This matrix features in two-point correlation
functions
in non-Hermitian systems,  e.g. in description of the
particle
escape from the scattering region (``norm leakage",
see Ref.~\onlinecite{SS}).

Following Ref.~\onlinecite{CM},  consider
two correlation functions:  a { diagonal} one
\begin{equation}\label{dia}
O({\cal E})=\Big\langle \frac{1}{N}\sum_{n}{\cal
O}_{nn}\,
\delta\left({\cal E}-{\cal
E}_k\right)\Big\rangle_{{\cal H}_N}
\end{equation}
and an  { off-diagonal} one
\begin{equation}\label{ndia}
O({\cal E}_1,{\cal E}_2)=
\Big\langle \frac{1}{N}\!\!\sum_{n\ne m}\!\!{\cal O}_{nm}\,\,
\delta\!\left({\cal E}_1\!-\!{\cal E}_n\right)\,
\delta\!\left({\cal E}_2\!-\!{\cal
E}_m\right)\Big\rangle_{{\cal H}_N}.
\end{equation}
Here $\langle\cdots\rangle_{{\cal
H}_N}$ stands for 
an ensemble average over ${\cal H}_N$.
The correlation functions (\ref{dia},\ref{ndia}) 
characterise the average non-orthogonality of
eigenvectors
corresponding to resonances whose positions in the
complex
plane are  close to the complex energies ${\cal E}$,
and ${\cal E}_1,{\cal E}_2$.
Here $\delta({\cal E})$ stands for a two-dimensional
$\delta-$function of the complex
variable $\cal E$. 

In the context of lasing media, the diagonal correlator (\ref{dia})
characterises average excess noise factors (Petermann factors), and the off-diagonal correlator
(\ref{ndia}) describes average cross correlations between
thermal or quantum noise emitted into different
eigenmodes \cite{Pet}.
Note that for any ensemble with orthogonal eigenvectors and
complex
eigenvalues ${\cal E}$ (for {normal}
matrices), $O({\cal E})$ is
equal to the mean density 
of
complex eigenvalues, and the off-diagonal
correlator vanishes: $O({\cal E}_1,{\cal E}_2)\equiv 0$.

Both diagonal and off-diagonal eigenvector correlators
were
introduced and calculated
for the case of Ginibre's ensemble of non-Hermitian
matrices in Ref.~\onlinecite{CM}. For the ensemble ${\cal H}_N$
 pertinent to chaotic 
scattering, both types of eigenvector correlators were
found recently
for the regime of very strongly overlapping resonances
when widths typically much exceed the mean
separation\cite{Janik,MS}. 
Physically this regime corresponds to a situation
where the scattering system 
is coupled to the continuum via a large number $M\gg
1$ 
of open channels\cite{FS}. In this case the
self-consistent 
Born approximation is adequate\cite{CM,chalker,MS,Janik},
a perturbative approximation valid in the limit of
large $N$, large $M$, and $|{\cal E}_1-{\cal E}_2|\neq
0$, provided ${\cal E}_{1},{\cal E}_{2}$ are well
inside the support of the spectrum.
A non-perturbative expression for the diagonal
correlator
$O(z)$ valid for any number of open channels was
obtained in  Ref.~\onlinecite{Schom} by employing a heuristic 
analytic continuation procedure.
For the case of the resonance widths, this 
heuristic scheme is known to reproduce the exact
expression
\cite{FS}. It is thus natural to expect that 
this procedure is adequate in the case of
eigenvector statistics, too, although
this remains to be proven.
 
No non-perturbative results for the off-diagonal
eigenvalue correlator $O({\cal E}_1,{\cal E}_2)$
have so far been reported, to the best of our
knowledge.

In the present paper we provide exact non-perturbative
expressions for both diagonal and off-diagonal 
eigenvector correlators valid for
the case of a system with broken time-reversal
invariance (anti-unitary symmetry)
 coupled to continuum via a single open channel
($M=1$) with 
coupling strength $\gamma$. The single-channel case
describes pure resonant chaotic reflection. 
This case is more amenable
to analytical treatment than a general case ($M>1$),
combining both reflection and transmission phenomena.
Understanding the single-channel case should be considered as a
useful
 step towards a more complete picture\cite{Izr}.  

Our result for the diagonal correlator is
\begin{eqnarray}\label{res1}
O({\cal E})=\nu e^{-4\pi g Y/\Delta}\frac{d}{dY}
\left\{ e^{2\pi g Y/\Delta} 
\frac{\sinh{(2\pi Y/\Delta)}}{2\pi Y/\Delta}\right\} 
\end{eqnarray}
where ${\cal E}=E-iY$, $\nu\equiv\nu(E)\,,\,
\Delta=\Delta(E)$
and $g=(\gamma+\gamma^{-1})/(2\pi\nu)$ is the
effective
(renormalised) coupling strength.
The result for $O({\cal E})$ agrees with one reported
in Ref.~\onlinecite{Schom} confirming the validity of the analytical
continuation
scheme used there. For the sake of comparison we
present here also the
expression for the single-channel resonance density
defined as $d({\cal E})=\left\langle N^{-1}\sum_k
\delta\left({\cal E}-{\cal
E}_k\right)\right\rangle_{{\cal H}_N}$ and given
by\cite{FS}:
\begin{eqnarray}\label{resden}
d({\cal E})=-\nu \frac{d}{dY}
\left\{ e^{-2\pi g Y/\Delta} 
\frac{\sinh{(2\pi Y/\Delta)}}{2\pi Y/\Delta}\right\} \,.
\end{eqnarray}%
\begin{figure}
\psfrag{XLABEL}{\mbox{}\hspace*{0.25cm}$Y$}
\psfrag{TAG1}{$O({\cal E})$}
\psfrag{TAG2}{$d({\cal E})$}
\centerline{\includegraphics[clip,width=5.5cm]{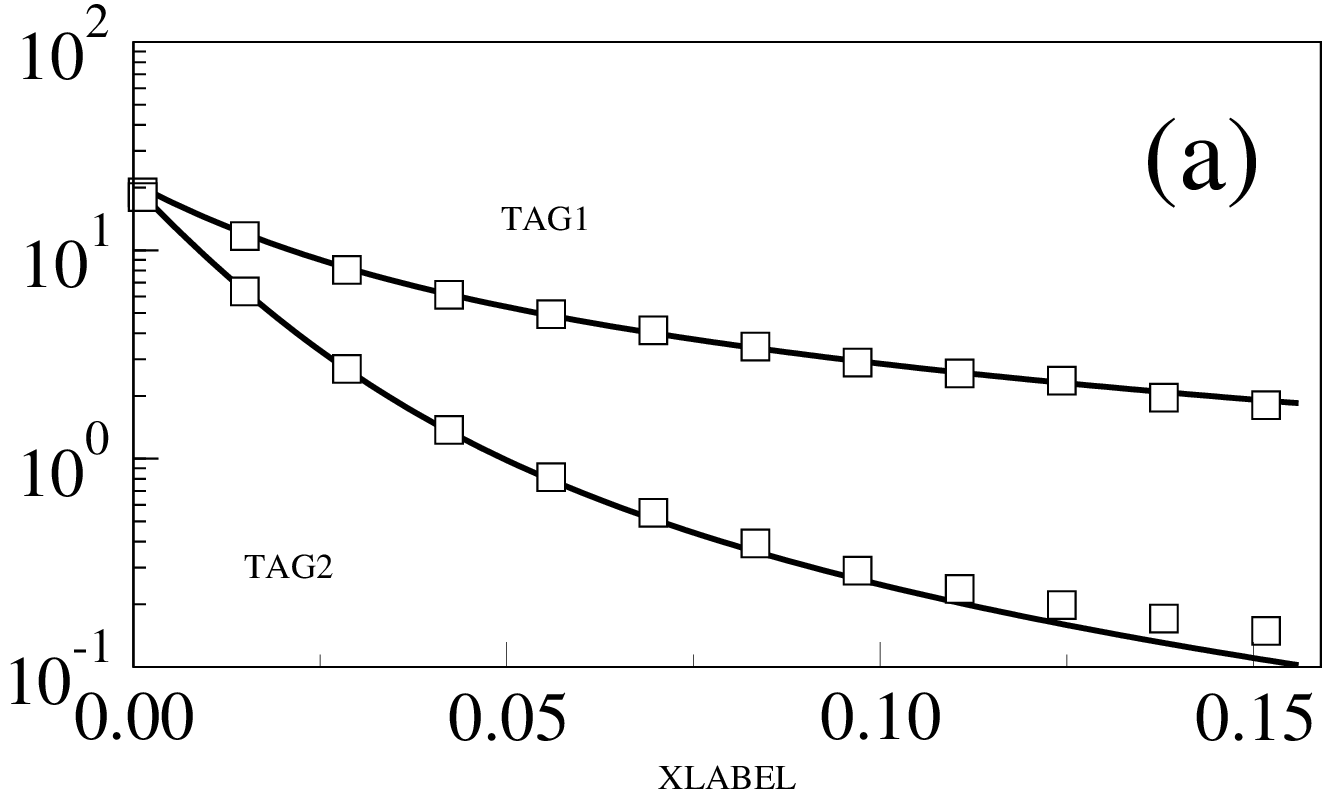}}
\mbox{}\\[-2mm]
\psfrag{XLABEL}{\mbox{}\hspace*{1.25cm}$L_y$}
\psfrag{TAG1}{$O_1(L_x,L_y)$}
\psfrag{TAG2}{$\langle n(L_x,L_y)\rangle$}
\centerline{\includegraphics[clip,width=5.5cm]{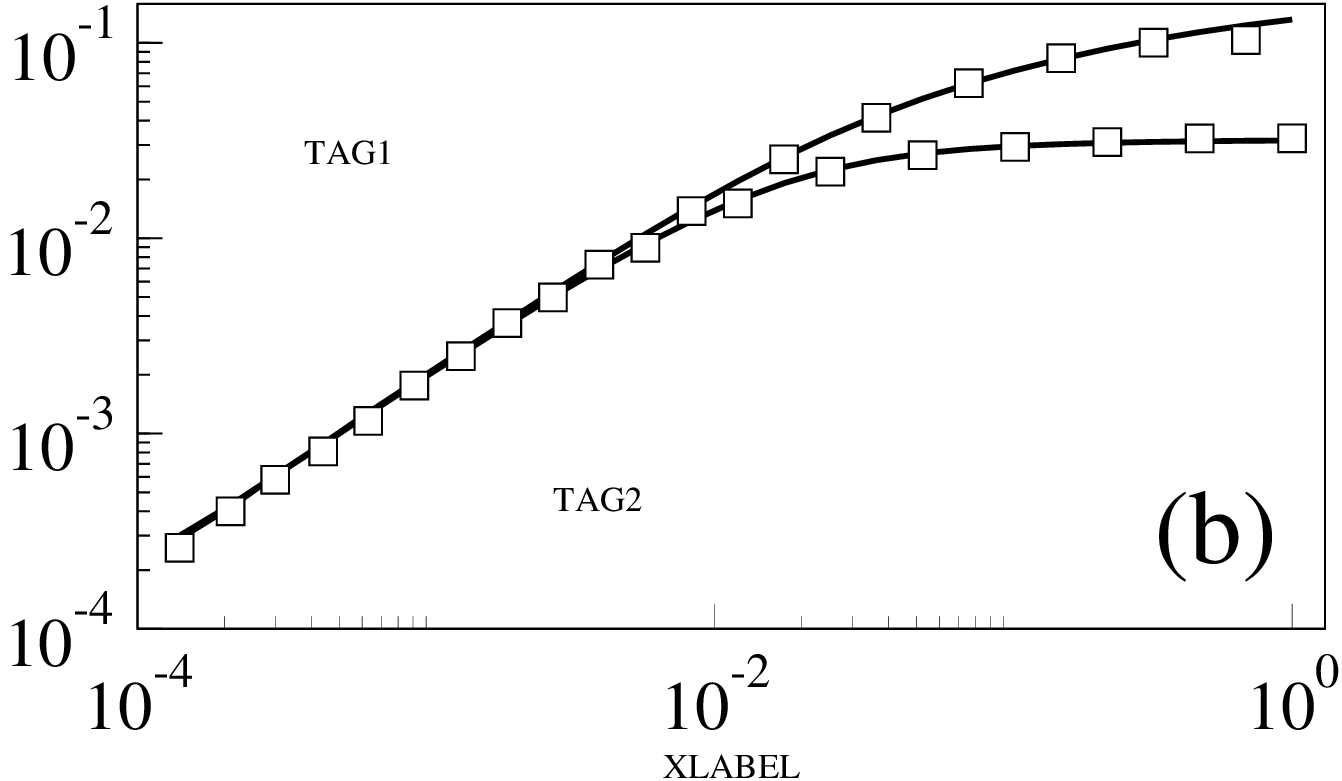}}
\mbox{}\\[-2mm]
\psfrag{XLABEL}{\mbox{}\hspace*{15mm}$L_y$} 
\psfrag{TAG1}{\mbox{}\hspace*{-1mm}$(-1)\!\times\!O_2(L_x,L_y)$}
\centerline{\includegraphics[clip,width=5.5cm]{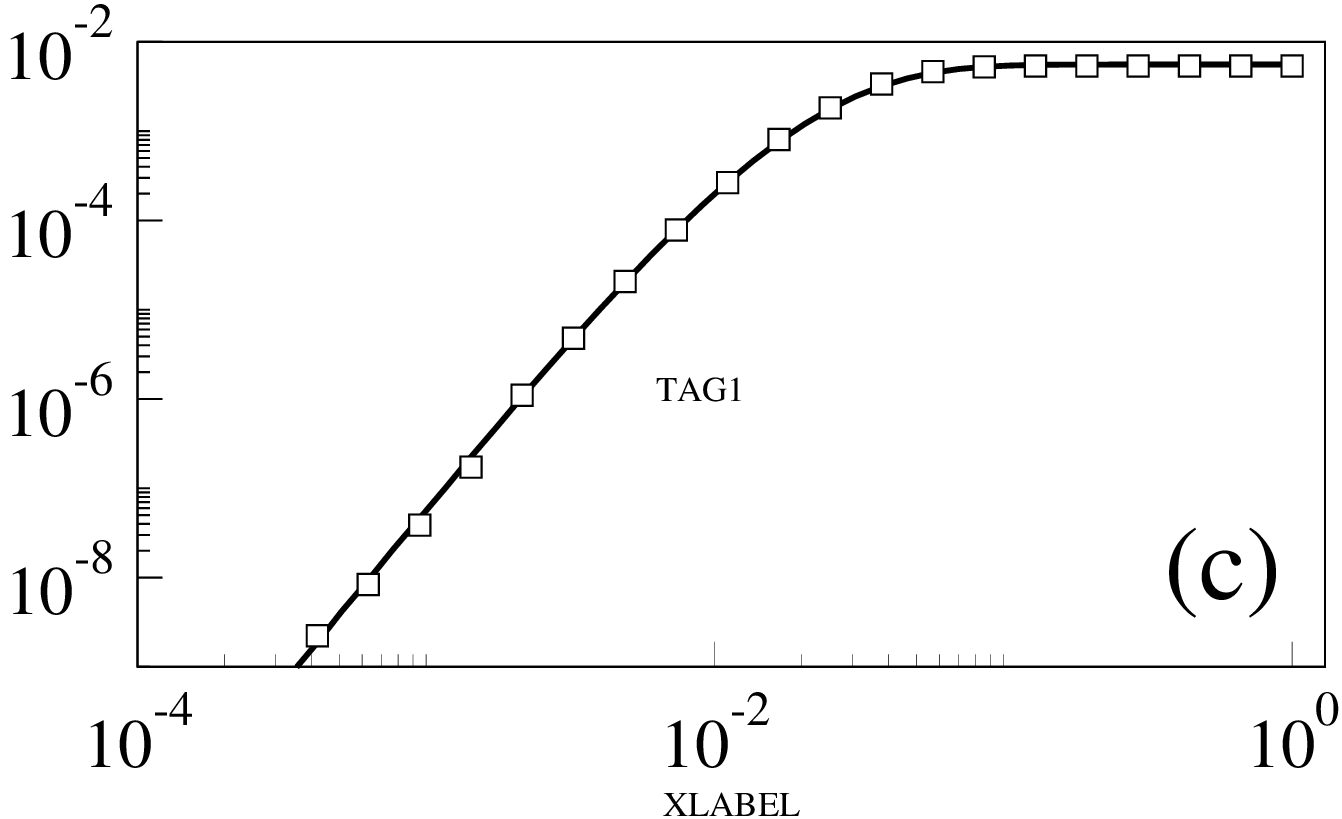}}
\mbox{}\\[-6mm]
\caption{\label{fig:diag_av}  Numerical 
($\protect\square$) and analytical results (solid line) for  (a)
$d({\cal E})$ and $O({\cal E})$ as a function of $Y$
for $N=32$, and $\gamma = 0.9$; 
(b) $\langle n(L_x,L_y)\rangle$
and $O_1(L_x,L_y)$
for $N=32$, $L_x=0.1$, and $\gamma = 0.9$;
(c)   $O_2(L_x,L_y)$
for $N=32$, $L_x=0.1$, and $\gamma = 0.9$.  }%
\end{figure}
\mbox{}\noindent We have 
compared these analytical expressions, valid in the
limit $N\rightarrow\infty$, with direct numerical
diagonalisations of finite-dimensional
matrices ${\cal H}_N$, see Fig.~\ref{fig:diag_av}(a). 
This is of interest since empirically, 
the ensemble average $\langle \cdots\rangle_{{\cal H}_N}$
is usually replaced by an energy average over
several spectral windows, each of which may typically contain
of the order of 10 or 100 resonances, corresponding
to a finite value of $N$. We observe that the analytical results
describe the numerical data well, except
for small deviations at large values of $Y$.
Numerically it is easier to compute smoothed
averages, such as the mean number of eigenvalues
$\langle n(L_x,L_y)\rangle$ inside 
a rectangular domain
\begin{equation}\label{area}
A=\left\{\begin{array}{l}-L_x/2\le \mbox{Re}\,{\cal
E}\le L_x/2\\[0.2cm]
\mbox{}\hspace*{9mm}  0\le \mbox{Im}\,{\cal E}\le L_y
\end{array}\right.
\end{equation}
in the complex plane.  This quantity can be obtained
from the mean density $d({\cal E})$
 by integration over $A$.
Similarly one can define the function $O_1(L_x,L_y)$
as the integral
of the diagonal correlator $O({\cal E})$ over the same
domain,
obtaining $O_1(L_x,L_y) = \langle N^{-1}\sum_{{\cal E}_k\in A}{\cal
O}_{kk}\rangle$. 
Numerical versus analytical 
results for these two quantities are plotted in
Fig.~\ref{fig:diag_av}(b).

For the off-diagonal correlator $O({\cal E}_1,{\cal
E}_2)$ 
we obtain
\begin{eqnarray}\label{res2}
&&O({\cal E}_1,{\cal
E}_2)=N({\pi\nu}/{\Delta})^2e^{-2\pi g
(Y_1+Y_2)/\Delta}\\
\nonumber &\times&
\int_{-1}^{1}d\lambda_1\int_{-1}^{1}
d\lambda_2(g+\lambda_1)(g+\lambda_2)e^{i\pi\Omega(\lambda_1+\lambda_2)/\Delta}
\\ \nonumber &\times&
e^{-\pi Y_2 (\lambda_1-\lambda_2)/\Delta}
\left[ e^{\pi Y_1(\lambda_1-\lambda_2)/\Delta} 
-e^{-\pi Y_1(\lambda_1-\lambda_2)/\Delta}\right] 
\end{eqnarray}
where $\mbox{Re}\,{\cal E}_{1,2} = E_{1,2}=E\mp \Omega$ and it is assumed that 
$\Omega\sim \Delta$. We have also calculated
the corresponding smoothed average 
$O_2(L_x,L_y) = \langle N^{-1}\sum_{{\cal E}_m\neq{\cal E}_n\in
A}{\cal O}_{mn}\rangle$,
by integrating ${\cal E}_1$ and ${\cal E}_2$ in
(\ref{res2}) 
over the domain $A$. In Fig.~\ref{fig:diag_av}(c) we
compare this result (valid in the limit of
$N\rightarrow\infty$)
with those of numerical diagonalisations of finite
matrices; the agreement is good already for $N=32$.
\begin{figure}
\psfrag{XLABEL}{\mbox{}\hspace*{2mm}$x$}
\psfrag{YLABEL}{$f(x)$}
\centerline{\includegraphics[clip,width=6cm]{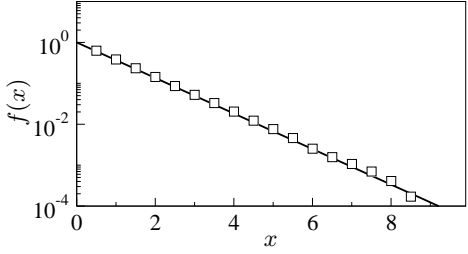}}
\mbox{}\\[-6mm]
\caption{\label{fig:hist} Numerical ($\square$) 
and analytical results (solid line) for
the distribution of the narrowest resonance for
$\beta=2$, and $\gamma = 0.1$, $W=0.2$, and $N=128$.
Here $x = \pi g n\,\Gamma/\Delta$.  }
\end{figure}

We have also found a way to calculate exactly the
distribution 
$f(\Gamma)$ of the width of the most narrow resonance
among 
 those falling in a window 
$[E-W/2,E+W/2]$ in the vicinity
of a given point $E$ in the spectrum. Assuming that
the mean number
$n=W/\Delta$ of resonances is large ($n \gg 1$), but
still $W\ll 1$ to
preserve spectral locality (the density of states
should not change significantly across the spectral window):
\begin{equation}\label{res3} 
f(\Gamma)=\frac{\pi g n}{\Delta}\, e^{-\pi g n
\Gamma/\Delta}\,.
\end{equation}
This distribution is of
great interest in the theory of random
lasing\cite{Schom}.
The functional form of the distribution was
found in Ref.~\onlinecite{Schom} by employing plausible
qualitative arguments yielding Eq.~(\ref{res3}), but with
renormalised effective
coupling $g$ replaced by its ``weak coupling" limit
$\gamma/2\pi{\nu}$. 
We see that the difference with
exact formula amounts to the factor $2$ in the
exponent for the case
of perfect coupling $\gamma=1$.  In Fig.~\ref{fig:hist}, the result
(\ref{res3}) is compared to results of numerical
diagonalisations
for $N=128$ and $W=0.2\Delta$, corresponding
to $n \approx 8.15$.

In the remainder of this article, we outline the
derivation of the 
results (\ref{res1}),(\ref{res2}),(\ref{res3}). The
main idea is to use that fact that the complex
eigenvalues (resonances) ${\cal E}_k$ are poles of the
$M\times M$ scattering matrix $\hat{S}({\cal E})$
in the complex energy plane. Using the
standard expression for the scattering matrix in terms
of the
non-Hermitian Hamiltonian ${\cal H}_N$ (see
e.g. Ref.~\onlinecite{FS}) the residues corresponding to these poles can
be found and
after some
algebraic manipulations we arrive at the following
relation:
\begin{eqnarray}\label{residues}
 &&\mbox{Tr}\left\{\mbox{Res}
\left[\hat{S}({\cal E})\right]_{{\cal E}={\cal E}_{n}}
\mbox{Res}\left[\hat{S}^{\dagger}(\tilde{{\cal
E}}^{*})
\right]_{\tilde{{\cal E}}^{*}={\cal
E}^{*}_{m}}\right\}
\\ \nonumber
&&\mbox{}\hspace*{-1cm}=\left( {\cal E}_m^*-{\cal
E}_n\right)
\left({\cal E}_n-{\cal E}^*_m\right)\,\, 
{\cal O}_{mn}\,.
\end{eqnarray}
This relation is valid for arbitrary $M$, but for
$M>1$ it appears
to be of no obvious utility, due to difficulties in
evaluating the ensemble average of the trace of the
residues on the left-hand side. However for the
case of one single open
channel the scattering matrix can be written as
\begin{equation}\label{1ch}
S({\cal E})=\prod_{k=1}^N\frac{{\cal E}-{\cal E}^*_k}
{{\cal E}-{\cal E}_k}\quad,\quad
S^{\dagger}({\cal E})=\prod_{k=1}^N\frac{{\cal
E}^*-{\cal E}_k}
{{\cal E}^*-{\cal E}^*_k} 
\end{equation}
which follows, up to an irrelevant
``non-resonant" phase factor,
 from the requirement of $S-$ matrix 
analyticity in the upper half-plane and unitarity for
real energies.
Substituting Eq.~(\ref{1ch}) into Eq.~(\ref{residues})
yields the relation:
\begin{equation}\label{overlap}
{\cal O}_{mn}\!=\!\frac{\left({\cal E}_n\!\!-\!{\cal E}^*_n\right)
\left({\cal E}_m\!\!-\!{\cal E}^*_m\right)}
{\left({\cal E}_n\!\!-\!{\cal E}^*_m\right)^2}
\prod_{k\ne n}^N\frac{{\cal E}_n\!\!-\!{\cal E}^*_k}
{{\cal E}_n\!\!-\!{\cal E}_k}
\prod_{k\ne m}^N\frac{{\cal E}^*_m\!\!-\!{\cal E}_k}
{{\cal E}^*_m\!\!-\!{\cal E}^*_k}
\end{equation}
expressing the eigenvector overlap
matrix
in terms of complex eigenvalues ${\cal
E}_k$\cite{note}. 
This gives a possibility to find the diagonal and
off-diagonal correlators,
Eqs.~(\ref{dia},\ref{ndia}), by averaging ${\cal
O}_{mn}$ over known joint
probability density of complex eigenvalues \cite{FK}
for the
single-channel scattering system:
\begin{eqnarray}\label{jpd}
&&{\cal P}\left({\cal E}_1,...,{\cal E}_N\right)=
\frac{e^{\frac{-N}{2}\gamma^2}}{\gamma^{N-1}}
|\Delta\{{\cal E}_1,...,{\cal E}_N\}|^2\\ \nonumber
&\times& e^{-\frac{N}{4}\sum_{k}({\cal E}^2_k+{\cal
E}^{*2}_k)}
\delta\Big(\gamma-\sum_{k}\mbox{Im}{\cal E}_k\Big)\,.
\end{eqnarray}
Using this expression one may notice that 
\begin{eqnarray}\label{dia1}
&&O({\cal E})=
\frac{\tilde{\gamma_1}^{N-2}}{\gamma^{N-1}}
e^{-\frac{1}{2}
\left[N\gamma^2-(N-1)\tilde{\gamma_1}\right]}
e^{-\frac{N}{4}({\cal E}^2+{\cal E}^{*2})}\\
\nonumber &&\times 
\left\langle \det{\left({\cal E}-{\cal
H}^{\dagger}\right)
\left({\cal E}^*-{\cal
H}\right)}\right\rangle_{\tilde{\cal H}_{N-1}}
\end{eqnarray}
where $\tilde{\cal H}_{N-1}$ stands for the
non-Hermitian matrix ${\cal H}$ of the same type as ${\cal H}_N$
but of the lesser size $(N-1)\times(N-1)$, and with
coupling $\gamma$ replaced by 
a modified coupling
$\tilde{\gamma_1}=\gamma-\mbox{Im}\,{\cal E}$.
Analogously
\begin{eqnarray}\label{dia2}
&&O({\cal E}_1,{\cal
E}_2)=\frac{\tilde{\gamma_2}^{N-3}}{\gamma^{N-1}}
e^{-\frac{1}{2}\left[N\gamma^2-(N-2)\tilde{\gamma_2}\right]}\\
\nonumber &&\times 
e^{-\frac{N}{4}\sum_{n=1}^{2}({\cal E}_n^2+{\cal
E}_n^{*2})}
({\cal E}_1-{\cal E}_1^*)({\cal E}_2-{\cal E}_2^*)\\
\nonumber &&\times \left\langle \det{\left({\cal
E}_1-{\cal H}^{\dagger}\right)
\left({\cal E}_1^*-{\cal H}^{\dagger}\right)
\left({\cal E}_2-{\cal H}\right)
\left({\cal E}_2^*-{\cal H}\right)
}\right\rangle_{\tilde{\cal H}_{N-2}}
\end{eqnarray}
where $\tilde{\cal H}_{N-2}$ is of
the size $(N-2)\times(N-2)$, and with coupling
$\gamma$ replaced by 
a modified coupling
$\tilde{\gamma_2}=\gamma-\mbox{Im}\,{\cal E}_1
-\mbox{Im}\,{\cal E}_2$.

In this way the problem is reduced to calculating a
correlation function
of characteristic polynomials of large non-Hermitian
matrices. A closely related object was calculated in
Ref.~\onlinecite{FK}, and we can
adopt those methods to our case. The scaling limit
$N\gg 1$ such that 
$\mbox{Im}\,{\cal E}_{1,2}=\Gamma_{1,2}\sim 
2\Omega=\mbox{Re}\left(
{\cal E}_{1}-{\cal E}_{2}\right)\sim \Delta\propto
N^{-1}$
of the resulting expressions yields the formulas 
Eqs.~(\ref{res1})-(\ref{res2})
above.

Let us briefly comment on a way of calculating the
distribution
Eq.~(\ref{res3}) of the widths of the most narrow
resonance in a given window.
Instead of extracting such a quantity from the joint
probability density 
Eq.~(\ref{jpd}) we find it more convenient to 
consider
\begin{eqnarray}\label{jpd1}
&&{\cal P}\left(z_1,...,z_n\right) \propto
\frac{1}{T^{n-1}}
|\Delta\{z_1,...,z_n\}|^2\\ \nonumber
&&\times 
\delta\Big(1-T-\prod_{k=1}^n\mbox{Im}|z_k|^2\Big)
\end{eqnarray}
defined for complex variables $z_i=r_ie^{\theta_i}$ 
inside the unit circle: $r_i=|z_i|\le 1$.
For $0\le T\le 1$ this formula has interpretation of
the joint probability density of complex eigenvalues
$z_i$ for the ensemble of $n\times n$  subunitary matrices
and is a very natural ``circular" analogue of
Eq.~(\ref{jpd}).
The similarity is in no way a superficial one, but
rather 
has deep roots in the theory of scattering\cite{F}.
The parameter $T$
controls the deviation of the corresponding matrices
from unitarity, 
much in the same way as the parameter $\gamma$
controls the deviation
of ${\cal H}_N$ from Hermiticity. More precisely, $T$
should be
associated with the renormalised coupling constant $g$
via the relation $g={2}/{T}-1$.
In the limit $n\gg1$ the eigenvalues $z_k$ are
situated in a narrow
vicinity of the unit circle. Their statistics is
shown\cite{F} to be
indistinguishable from that of the complex
eigenenergies ${\cal E}$,
when the latter considered {\it locally}, 
i.e. on the distances comparable with the mean spacing
$\Delta$. 
In particular, the distances $1-r_i$ from the unit
circle 
should be interpreted as the widths of the resonances.

The form of
Eq.~(\ref{jpd1}) allows one to integrate out the
phases $\theta_i$
by noticing that:
\begin{equation}
\int_0^{2\pi}\!\!\frac{d\theta_1}{2\pi}\ldots
\int_0^{2\pi}\!\!\frac{d\theta_n}{2\pi}
\!\prod_{k<j}\!|r_k e^{i\theta_k}\!-\!r_j e^{i\theta_j}|^2
\!=\!\!\sum_{\{\alpha\}}r_1^{2\alpha_1}\!\ldots
r_n^{2\alpha_n}
\end{equation}
where the summation  goes over all possible 
permutations $\{\alpha\}=(\alpha_1,...,\alpha_n)$ of
the set
$1,...,n$ (in fact in the right-hand side we deal with
the object known as ``permanent", see e.g. Ref.~\onlinecite{per}).
In this way we arrive at a joint
probability density of the radial coordinates only,
and
the distribution Eq.~(\ref{res3}) follows after a
number of integrations and the limiting procedure $n\gg1$. 

In conclusion, we presented a detailed analytical and
numerical 
investigation of statistics of resonances and
associated bi-
orthogonal eigenfunctions in a random matrix model of
single channel chaotic
 scattering with broken time-reversal invariance. 
Among challenging problems deserving future research 
we would like to mention extending our results to the
case of more than one channel 
and to time-reversal invariant systems\cite{FTS1}, as
well as the problem of understanding
fluctuations of the non-orthogonality overlap matrix
${\cal O}_{mn}$. 

{\em Acknowledgements.} Financial support by  EPRSC
Research Grant GR/13838/01
(YVF) and by Vetenskapsr\aa{}det (BM) is acknowledged
with thanks.


\begin{thebibliography}{10}
\bibitem{CM} J. T. Chalker and B. Mehlig, {\it
Phys. Rev. Lett.} {\bf 81} (1998), 3367; 
B. Mehlig and J. T. Chalker, {\it J. Math. Phys.} {\bf 41}
(2000), 3233
\bibitem{Janik} R. A. Janik {\em et al.},  {\it Phys. Rev. E}
{\bf 60} (1999), 2699
\bibitem{Schom} H. Schomerus {\em et al.}, {\it Physica A}
{\bf 278} (2000),
469; K. Frahm {\em et al.}, {\it Europh. Lett.} {\bf 49}
(2000), 48;
M. Patra {\em et al.}, {\it Phys. Rev. A} {\bf 61}, (2000)
23810
\bibitem{Greb} S. Yu. Grebenschikov, {\it J. Phys. Chem.
A} {\bf 104} (2000) 10409 
\bibitem{MS} B. Mehlig and M. Santer, {\it Phys. Rev. E}
{\bf 63} (2001), art. no. 020105
\bibitem{chalker}  J. T. Chalker and Z. J. Wang, {\it
Phys. Rev. Lett.} {\bf 79} (1997), 1797
\bibitem{Haake} G. Hackenbroich {\em et al.},
arXiv:quant-ph/0111156; C. Viviescas and 
G. Hackenbroich arXiv:quant-ph/0203122 
\bibitem{FS} Y. V. Fyodorov and H.-J. Sommers,
{\it J. Math. Phys.} {\bf 38}, (1997), 1918 and {\it
JETP Lett.} {\bf 63} (1996), 1026 
\bibitem{mehta}M.~L. Mehta, {\em
Random Matrices} (Academic Press,
Boston, 1990).
\bibitem{FK} Y. V. Fyodorov and B. A. Khoruzhenko,
{\it Phys. Rev. Lett.} {\bf 83} (1999), 65
\bibitem{Kottos} T. Kottos and U. Smilansky, {\it
Phys. Rev. Lett.} 
{\bf 85} (2000), 968; H. Ishio, {\it Phys. Rev. E} {\bf
62} (2000), 3035 
\bibitem{Kol} M. Gl\"{u}ck, A. R. Kolovsky and
H. J. Korsch, 
{\it Phys. Rev. E} {\bf 60} (1999), 247
\bibitem{SS} D. V. Savin and V. V. Sokolov, {\it
Phys. Rev. E} {\bf 56}
(1997), 4911
\bibitem{Pet}K. Petermann {\it IEEE J. Quantum Electron} {\bf 15} (1979),  566; H.A.Kaus and S. Kawakami {\it IEEE J. Quantum Electron} {\bf 21}
 (1985),  63; A. E. Siegman {\it Phys. Rev. A} {\bf 39} (1989) 1253 
 and {\it ibid.}  {\bf 39} (1989), 1264
\bibitem{Izr} F. Izrailev {\em et al.}, {\it Phys. Rev. E} {\bf 49} (1994), 130 
\bibitem{note}  Similar ideas were used in Refs.~\onlinecite{Greb} and \onlinecite{Izr}.
\bibitem{F} Y. V. Fyodorov in: ``Disordered and Complex
Systems", edited by P.Sollich et al.,
 AIP Conference Proceedings v. 553, p. 191 (Melville
NY, 2001);
Y.V.Fyodorov and H.-J.Sommers, {\it JETP Letters}, {\bf
72} (2000), 422 
\bibitem{FTS1}H.-J. Sommers {\em et al.}, {\it J.  Phys. A} {\bf 32} (1999) L77 
\bibitem{per} M.L. Mehta, ``Matrix Theory: Selected
topics and useful
results" (Dehli:Hindustan Publishing Corporation,1989)
p.14 
\end{thebibliography}
\end{document}